\documentclass[epsf,12pt]{article}
\usepackage{graphicx,caption2}
\begin{document}

\def\be{\begin{equation}}
\def\ee{\end{equation}}
\def\bea{\begin{eqnarray}}
\def\eea{\end{eqnarray}}
\def\tr{ \mathop{\rm tr}}
\def\atanh{\mathop{\rm atanh}}
\def\Tr{\mathop{\rm Tr}}
\def \ln {{\rm ln}\, }
\def \cotg {\rm cotg }
\addtolength{\arraycolsep}{-\arraycolsep}


\begin{center}
{\Large\bf Yang-Lee Zeros of the Ising model on Random 
          Graphs of Non Planar Topology}
\vskip 1.1cm
{\bf Luiz C. de Albuquerque\footnote{claudio@fma.if.usp.br}}
\vskip 0.1cm
{\it Departamento de F\'{\i}sica Matem\'atica, Instituto de F\'{\i}sica -
         USP \\
         C.P. 66318, S\~ao Paulo, SP, Brazil}
\vskip 0.3cm
{\bf Nelson A. Alves\footnote{alves@quark.ffclrp.usp.br}}
\vskip 0.1cm
{\it Departamento de F\'{\i}sica e Matem\'atica, FFCLRP - USP \\
         Av. Bandeirantes 3900. CEP 014040-901 Ribeir\~ao Preto, SP, Brazil}
\vskip 0.3cm
{\bf D. Dalmazi\footnote{On leave from UNESP (Guaratinguet\'{a}) -
                 dalmazi@tonic.physics.sunysb.edu}}
\vskip 0.1cm
{\it Department of Physics and Astronomy - SUNY \\
     Stony Brook,  NY 11794, USA} 
\vskip 0.3cm
\today
\vskip 0.4cm
\end{center}

\begin{abstract}
  We obtain in a closed form the $1/N^2$ contribution to the
free energy of the two Hermitian $N\times N$ random matrix model
with non symmetric quartic potential. 
  From this result, we calculate numerically the Yang-Lee zeros 
of the 2D Ising model on dynamical random graphs 
with the topology of a torus up to $n=16$ vertices. 
  They are found to be located on the unit circle on the complex
fugacity plane. 
In order to include contributions of even higher topologies
we calculated 
analytically the nonperturbative (sum over all genus) partition function 
of the model
$Z_n = \sum_{h=0}^{\infty} \frac{Z_n^{(h)}}{N^{2h}} $
for the special cases of $\, N=1,2$ and 
graphs with $n\le 20 $ vertices.
Once again  the Yang-Lee zeros are shown numerically  
to lie on the unit circle on the complex fugacity plane.
Our results thus generalize previous numerical 
results on random graphs 
by going beyond the planar approximation
and strongly indicate that there might be
a generalization of the Lee-Yang circle theorem
for dynamical random graphs. 
\vskip 0.1cm
{\it Keywords:} Yang-Lee zeros, Lee-Yang theorem,
Ising model, random matrix, random surfaces , 2D gravity .

{\it PACS-No.:} 05.50.+q, 05.70.Fh, 64.60.Cn
\vskip 0.1cm
\end{abstract}


\section{Introduction}
\indent

  Yang and Lee have established long ago \cite{a1} that the statistical
theory of phase transitions is connected with the distribution of the
zeros of the grand partition function on the complex fugacity plane.
  They have proved that, in the thermodynamic limit, 
those zeros circumscribe closed regions on the fugacity plane
where physical quantities remanin  analytic
, thus 
defining different phases of the system.  
Although complex values for the fugacity are non physical, 
the physical features of the phase transition can be obtained from 
the study of the distribution of the zeros which increase in number
and tend to pinch the positive real axis 
in the termodynamic limit .

 In \cite{a2} Lee and Yang have proved that the complex zeros of the
partition function of the ferromagnetic Ising
model in a complex magnetic field $H$ lie on the unit circle
in the complex $y$-plane $(y = e^{-2\beta H})$. 
  This result,
known as the circle theorem, makes no assumptions about the 
details of 
the lattice like its topology , coordination number , etc.  
Their starting point is the partition function 
of the model on a given static lattice which we 
can assume to be , for instance , a graph $G_n$ with $n$ vertices. 
The corresponding partition function
is defined as
\be
Z(G_n)\,=\, \sum_{\left\{\sigma_i\right\}}  
             e^{\left(\frac{\beta}{2} \sum_{i,j}G_{ij}
\sigma_i\sigma_j \, + \, H\sum_{i=1}^n \sigma_i \right)}\, ,  \label{eq:r1}
\ee 
where $\beta=1/T$ and $G_{ij}=1$ for nearest neighbors and
$G_{ij}=0$ otherwise. We have absorbed the factor $\beta$ in the 
definition of the magnetic field.
  As usual the configuration of the system  is determined by
specifying $\sigma_i=\pm 1$ on each vertex of $ \, G_n \, $. 
  Henceforth we assume that
$G_n$ is a closed (no external legs) graph  
 with $n$ four-legged vertices. It is important to emphasize 
that  $ \, G_n \, $ does not need to be regular and one has 
$G_{ii}=1$ for a link that connects a vertex to itself
  Defining $\, d\, $
as the number of spins down we can  rewrite $Z(G_n)$ as being 
proportional to a polynomial in the fugacity $y$,
\be
Z(G_n)\,=\, (cy^{1/2})^{-n}\,\sum_{d=0}^{n}\, P_d\, y^d\, ,    \label{eq:r2}
\ee
where $c=e^{-2\beta}$ and each of the coefficients 
$P_d$ corresponds to a partition function of
the 2D Ising model on $G_n$ without magnetic field $(H=0)$ with
$d$ spins down. Therefore they are all positive real and satisfy
$P_d=P_{n-d}$. Based on this and other properties of $P_d$ Lee-Yang
have proved their theorem \cite{a2} assuring that all zeros of $Z(G_n)$
lie on the unit circle : $\vert y_k\vert=1 , \, (k=1,2,\cdots ,n) $.

  Following Kazakov and Boulatov \cite{a3,a4} we can define the Ising model
on dynamical random surfaces (2D Gravity) by treating the lattice
$G_n$ itself as a degree of freedom. Its partition function is 
obtained from the usual Ising model by summing also over all $G_n$
with $\, n \, $ vertices:
\be
Z_n\,=\, \sum_{\left\{G_n\right\}} Z(G_n) \, =  \,
         \sum_{i} Z(G_n^{(i)})  \,.                      \label{eq:r5}
\ee 
Here $i$ labels the different  graphs 
with  $\, n \, $ vertices .

 The above sum is not arbitrary. 
 In our case, where we consider four-legged vertices,
the weights of the graphs are the corresponding 
combinatorial factors of a $\phi^4$ field theory. 
In this case the model of [3,4] is known to have a third
order phase transition at the critical
temperature $\beta = log 2 \, $ from a
disordered ( high temperature ) to an ordered 
(low temperature ) phase.
Clearly
$Z_n$ has the same form of eq.(\ref{eq:r2})
but with new coefficients 
${\tilde P}_d=\sum_{i} P_d(G_n^{(i)})$\, 
Although the property ${\tilde P}_d={\tilde P}_{n-d}$ 
is still satisfied , this  
is not a sufficient condition for the 
Lee-Yang circle theorem  
to hold . In general the zeros of linear
combinations of polynomials have a complicated
relation to the zeros of the original basic polynomials. However , quite
surprisingly it has been observed numerically (see \cite{a5,a6}) for
dynamical planar graphs (spherical topology) wiht $1\le n\le 14$ vertices
that the Yang-Lee zeros are located on the unit circle.
It is instructive to illustrate this point for 
planar graphs with a  low number of vertices $( n=2,3,4 )$
where an analytic analysis is simpler.
From eq.(\ref{eq:r1}) we easily obtain the partition functions given
in Fig. 1 . The overall factor $(cy^{1/2})^{-n}$ of
a $G_n$ graph can be canceled by adding a constant to the energy in 
eq.(\ref{eq:r1}).
\begin{figure}[ht]
\renewcommand{\captionlabeldelim}{.~}
\renewcommand{\figurename}{Fig.}
\begin{center}
\begin{minipage}[t]{1.0\textwidth}
\centering
\includegraphics[width=9cm]{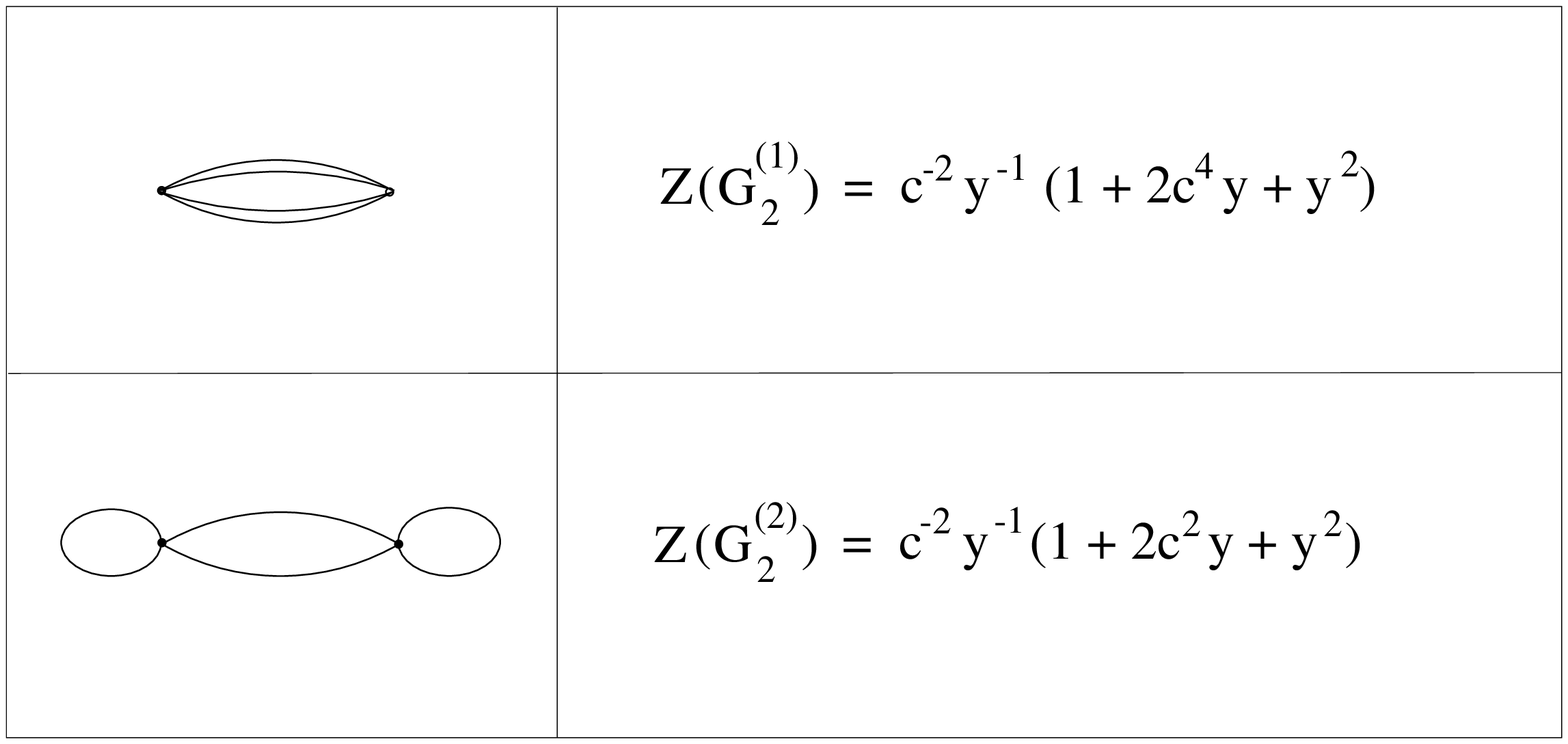}
\caption{Planar graphs and partition functions 
for $n=2$ vertices.}
\label{fig1}
\end{minipage}
\end{center}
\end{figure}
  Since $0\le c \le 1$, it is easy to verify that the Lee-Yang theorem holds
for $G_2^{(1)}$ and $G_2^{(2)}$ separately. This exemplify, in particular,
that the theorem does not depend on specific details of the graphs.
If we want to treat the graphs as an extra degree of freedom
we have to sum over $G_2^{(1)}$ and $G_2^{(2)}$ in this case.
Taking an arbitrary linear combination 
$Z_2(a,b) = a Z(G_2^{(1)}) + b Z(G_2^{(2)}) \,$ 
one can easily verify that  for 
$-2 < b/a < -2(\sqrt{2} - 1)$,\,
the zeros of $Z_2(a,b)$ will not belong to the unit circle. 
However, in the Kazakov and Boulatov's model
the combinatorial 
weights of $G_2^{(1)}$ and $G_2^{(2)}$ correspond
(see ref. \cite{a7} and the planar $Z_2$ displayed in [5])
to $(a,b)=(2,16)$, which brings the zeros to the unit circle. 

 In the next simplest case, $\, n=3 \, $, we have four planar graphs,
as shown in Fig. 2.
 Each graph gives rise to a polynomial of the form
\begin{figure}[ht]
\renewcommand{\captionlabeldelim}{.~}
\renewcommand{\figurename}{Fig.}
\begin{center}
\begin{minipage}[t]{1.0\textwidth}
\centering
\includegraphics[width=9cm]{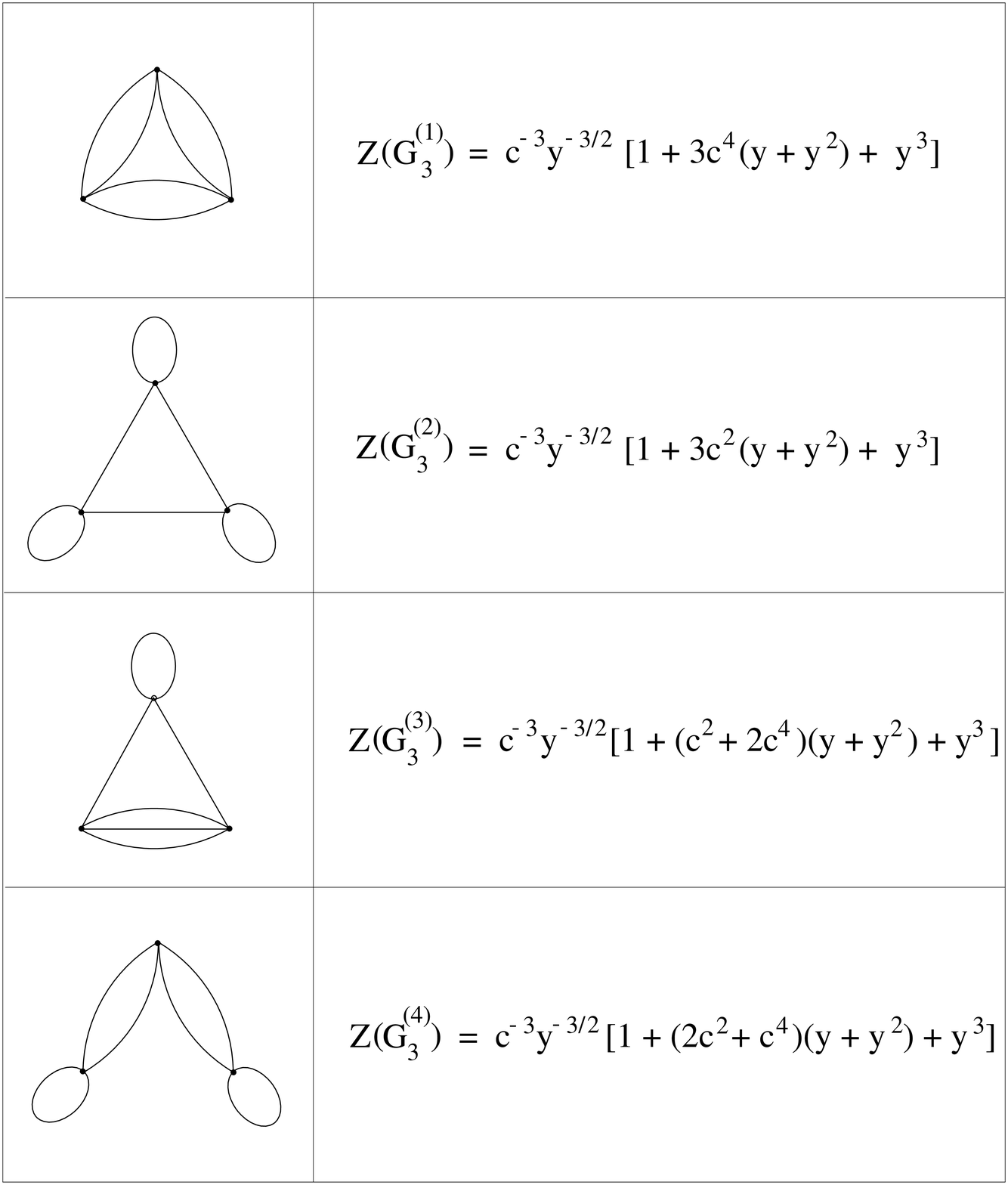}
\caption{Planar graphs  and partition functions for $n=3$ vertices.}
\label{fig2}
\end{minipage}
\end{center}
\end{figure}

\be
P_3^{(i)}\,=\, 1 \, +\, y^3 \, + \, 3 (y+y^2) a^{(i)} \, ,       \label{eq:s4} 
\ee
where the quantities $\, a^{(i)}\,$ are functions of
the temperature which satisfy 
\mbox{$0\le a^{(i)} \le 1$}, \,   $(i=1,2,3,4)\,$.
    It is easy to demonstrate that each $\, P_3^{(i)} \, $ has all its 
roots on the unit circle. 
Clearly, this is not true for a general  linear combination
$\, \sum_{i=1}^4 k_i\,P_3^{(i)} \,$. 
    Nevertheless, if $\, k_i \ge 0$,  the 
linear combination will have the same basic form of $\, P_3^{(i)} \,$
up to an overall constant,
and we will be back to the unit circle. 
Obviously, the combinatorial factors of the respective 
graphs (see ref. \cite{a7} and the planar $Z_3$ displayed in [5]): 
$\, k_i = (32/3,\, 256/3,\, 64,\, 128) $, belong to this subset.  

\begin{figure}[h]
\renewcommand{\captionlabeldelim}{a.~}
\renewcommand{\figurename}{Fig.}
\begin{center}
\begin{minipage}[t]{1.0\textwidth}
\centering
\includegraphics[width=9cm]{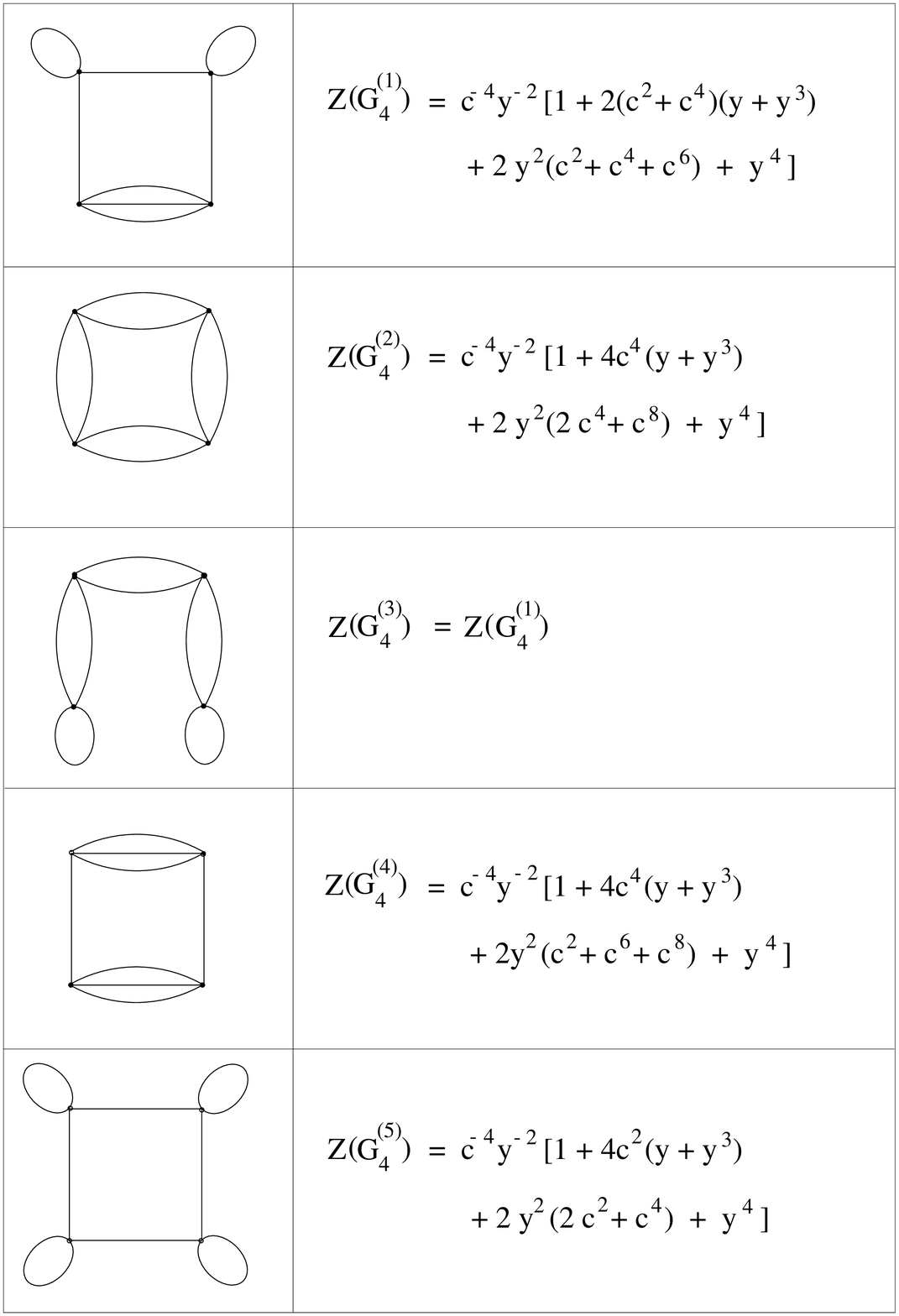}
\caption{First set of diagrams and partition functions for $n=4$ vertices.}
\label{fig3a}
\end{minipage}
\end{center}
\end{figure}

\setcounter{figure}{2}

\begin{figure}[h]
\renewcommand{\captionlabeldelim}{b.~}
\renewcommand{\figurename}{Fig.}
\begin{center}
\begin{minipage}[t]{1.0\textwidth}
\centering
\includegraphics[width=9cm]{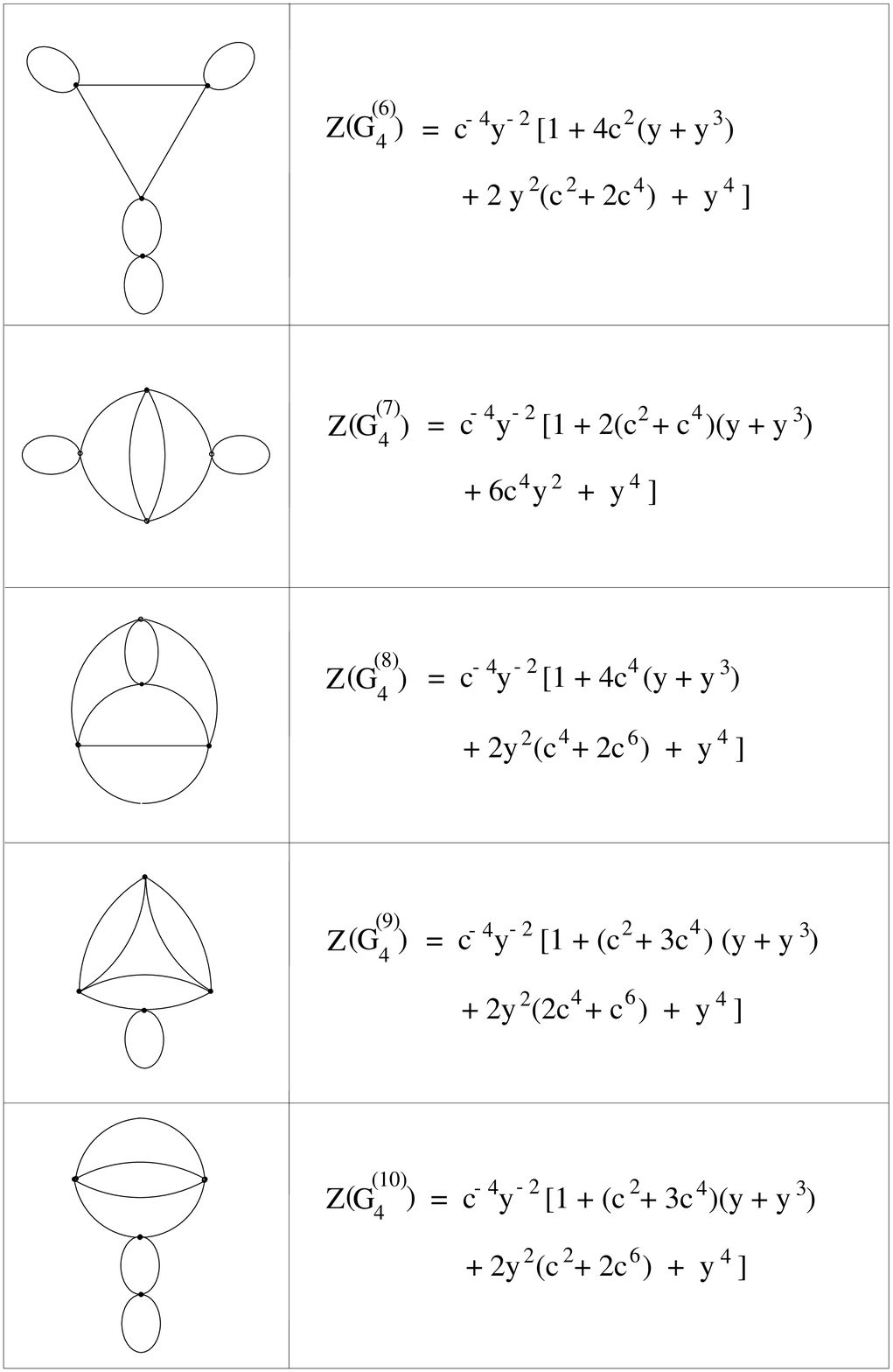}
\caption{Second set of diagrams and partition functions for $n=4$ vertices.}
\label{fig3b}
\end{minipage}
\end{center}
\end{figure}

 For $\, n=4 \, $ vertices there are ten planar graphs.
 In Fig. 3a and 3b we show the diagrams and the 
 corresponding partition functions.
 Each one corresponds to a polynomial, which can be written as:
\bea
P_4^{(i)} &=&  1 + y^4 + \left\lbrack a_1^{(i)}c^2 + (4-a_1^{(i)})c^4
  \right\rbrack (y+y^3)\, +\, 2\, \left\lbrack a_2^{(i)}c^2 
   + a_3^{(i)}c^4 \right\rbrack y^2                    \nonumber \\ 
& + & \, 2\, \left\lbrack (4 + a_1^{(i)}-3a_2^{(i)}-2a_3^{(i)}) c^6 + 
       (2a_2^{(i)}+a_3^{(i)}-1-a_1^{(i)})
       c^8\right\rbrack y^2   \, ,                         \label{eq:a}
\eea
where 
\bea
0\le a_1^{(i)}\le 4, &\makebox[0.7cm]{}&0\le a_2^{(i)},\, a_3^{(i)}\le 3,
                                                            \nonumber \\
4+a_1^{(i)}-3a_2^{(i)}-2a_3^{(i)}\ge 0, &\makebox[0.7cm]{}&
2a_2^{(i)} + a_3^{(i)} -1 - a_1^{(i)}\ge 0 \,.          \label{eq:b} 
\eea
  Analogous to the $\, n=3 \,$ case, each $\, P_4^{(i)} \,$, under 
conditions (\ref{eq:b}), possess only unit roots but that is
a property that cannot be extended to any  linear 
combination $\, \sum_{i=1}^{10}\, k_i\, P_4^{(i)} \, $. 
  However, once again, if $ \, k_i\ge 0 \, \, ( i=1,\cdots, 10 ) $ 
the conditions (\ref{eq:b}) and the polynomial form (\ref{eq:a}) 
would still hold for the linear combination and we end up with 
unit roots again. 
  Needless to say the combinatorial factors of the
graphs $G_4^{(i)}$, 
are all positive and therefore the roots will 
lie on the unit circle. 

It is remarkable
that even for larger $n$ the combinatorial factors 
of the graphs apparently, as our numerical results indicate, 
are such that the Yang-Lee zeros still lie on the unit circle.

For an arbitrary  
number of vertices
we were not able to make an analytical analysis of 
the location of the Yang-Lee zeros
for an arbitrary  finite temperature 
but  , as    one might 
expect from the results on static lattices,
the Yang-Lee zeros
of $Z_n$ should be equally distributed around the unit circle
at $T=0$ and coalesce at $y=-1$ as $T\to\infty\,$ .
As a self-consistency check we prove that this is
indeed the case directly from the two
Hermitian matrix model in the next section.
The key ingredient in those cases is the decoupling of the 
two Hermitian random matrix model into the one Hermitian
random matrix model.
   In section 3, using orthogonal polynomials, 
we obtain for arbitrary temperature the $1/N^2$
contribution to the free energy
of the two Hermitian matrix model.
   In section 4 we present the numerical results for the Yang-Lee zeros
on graphs of torus topology up to $n=16$ vertices. 
   In section 5 we calculate, for arbitrary temperature,
the non-perturbative partition function of the Ising model
on random lattices, including all topologies.
   For this end, we have to restrict ourselves to the 
cases of small matrices $N=1,2$ and graphs with
 $n\le 20$ vertices.
Numerical results show that even after summing over all topologies
we still have the Yang-Lee zeros on the unit circle.

\section{Free Energy at $T=0$ and $\, T\to \infty $ for arbitrary topology}
\indent 

By means of two Hermitian matrices $X$ and $Y$ of order $N\times N$, 
each one associated with a value of $\sigma_i =\mp 1 \, $, one can show that 
the free energy \cite{a3,a4}
\be
E(g,c,H)\, = -\frac{1}{N^2}\,{\rm log} 
\left( \frac{\int {\cal D}\mu\, e^{-\Tr \left\lbrack
 X^2+Y^2 \,-\, 2c X Y + \frac{g}{N} ( e^{H}X^4 + e^{-H}Y^4)
\right\rbrack }}
{\int {\cal D}\mu\, e^{-Tr \left\lbrack X^2+Y^2 \,-\, 2c X Y \right\rbrack }
}\right)                                                     \label{eq:r7}
\ee
is a generating function for partition functions $Z_n$
of the 2D Ising model in a constant magnetic field $H$ 
calculated on random graphs $G_n$ of $n$ vertices
of 4 links each
according to
\be
E(g,c,H)\, =\, -\sum_{n=1}^{\infty}\left\lbrack 
\frac{-gc}{(1-c^2)^2}\right\rbrack^n Z_n  \, .                   \label{eq:r8}
\ee
Notice that ${\cal D}\mu = {\cal D}X {\cal D}Y $ 
is the usual measure \cite{a9} for Hermitian matrices.
  The quadratic terms of the potential in eq.(\ref{eq:r7}) are 
responsible for the links  (propagators) 
between 
$<++>,<--> \, $ and $<+->\, $ 
while the quartic terms stand for the two different sites 
( vertices ) $\, (+) \, $ and $\, (-) \, $ 
of the lattices (graphs) each one with four links.
  This corresponds to a $\phi^4$ interaction field theory in which
the fields are represented by $N \times N$ matrices.

  Taking ratios of the propagators and comparing
with the Boltzmann weights of $Z(G_n)$ one identifies the temperature:
$ c\,=\, e^{-2\beta} $.
For the specific cases of $c=0\,$  $(T=0)\,$ and 
$\, c=1$  $\, (T\to \infty )$, the matrices in eq.(\ref{eq:r7}) 
decouple and the free energy $E(g,c,H)$ can be obtained 
for arbitrary topology in terms of
the one Hermitian matrix model free energy ($E_1$) as follows. 
Taking $c=0 \, (T=0) $ in eq.(\ref{eq:r7}) we have
\be
E(g,0,H)\,=\, E_1(\frac{ge^H}{4}) + E_1(\frac{ge^{-H}}{4})\, ,  \label{eq:r9} 
\ee
where 
\be 
E_1(g)\,=\, - \frac{1}{N^2}\, {\rm log}\,\left( \frac{\int{\cal D}X\, 
         e^{-\Tr \left(
   \frac{X^2}{2} +\frac{gX^4}{N} \right)}}{{\int{\cal D}X\, 
   e^{-\Tr \, \frac{X^2}{2}}}} \right) 
\, =\, - \sum_{n=1}^{\infty} (-g)^n \,a_n \, .    \label{eq:r10}
\ee
Plugging in eq.(\ref{eq:r9}), we obtain the following expansion
in the coupling $g$,
\be
E(g,0,H)\,=\, - \sum_{n=1}^{\infty} a_n (-g)^n (1 + y^n )\,
                 y^{-n/2}\, .                     \label{eq:r11}
\ee  
The coefficients $a_n$ on their turn have a topological expansion
\be
a_n \, =\, \sum_{h = 0}^{\infty}\,\frac{a_n^{(h)}}{N^{2h}}\, , \label{eq:r12} 
\ee
with $h=0,1,\cdots $, corresponding respectively to 
lattices with spherical topology, torus topology, etc. 
 The coefficients $a_n^{(h)}$ can be calculated iteratively using 
orthogonal polynomials for the one matrix 
model given in (10). From \cite{a7,a8} we have, for instance, the
sphere and torus contributions:
\bea
a_n^{(0)}\, & = & \,  \frac{3^n (2n-1)!}{n!(n+2)!} \\    \label{eq:r13}
a_n^{(1)}\, & = & \,  \frac{3^n}{24n}\left( 4^n - 
                     \frac{(2n)!}{(n!)^2}\right) \,.       \label{eq:r14}
\eea

\noindent 

  In the case $c\to 1 \, (T\to \infty) $, the
decoupling of the two matrices in eq.(\ref{eq:r7}) is less obvious. 
With the  redefinition 
$g={\overline g}(1-c^2)^2\,$ we obtain, in the limit $c\to 1$ , 
after a 
change of variables in eq.(\ref{eq:r7}),
\bea
E(g,c\to 1,H)\,&=&\,
  E_1\left( \frac{\overline g}{4} [y^{1/2} + y^{-1/2}] \right)  \nonumber \\
&=& - \lim_{c \rightarrow 1}\, \sum_{n=1}^{\infty}
\,a_n \frac{(-g)^n(1+y)^n y^{-n/2}}{(1-c^2)^{2n}} \,. \label{eq:r15}
\eea

 If we compare eq. (\ref{eq:r8}) with 
eqs.\,(\ref{eq:r11}) and\,(\ref{eq:r15}),
we obtain the partition functions
\be
Z_n \,(T=0) \, = \, a_n\,(1 + y^n)\,y^{-n/2}        \label{eq:r16}
\ee
and
\be
Z_n \,(T\to \infty ) \, =\, a_n\,(1 + y)^n\,y^{-n/2}\,. \label{eq:r17} 
\ee

Notice that we have dropped the factor 
$\left[ c/(1+c^2)^{2} \right]^n \, $
since it just amounts to a redefinition of the energy by a constant. 
  
   We conclude that for arbitrary number of sites the Yang-Lee zeros 
of the Ising model on dynamical random lattices of arbitrary 
topology\footnote{Though we have only given $a_n^{(0)}$ and $a_n^{(1)}$, 
                  the relations (\ref{eq:r16})
                  and (\ref{eq:r17}) hold for arbitrary topologies.} 
are homogeneously distributed around the unit circle   at $T=0$\,
$(y_k=e^{i \frac{\pi}{n}(2k-1)}$, \, $k=1,2,\cdots , n)\,$
and coalesce at the point $y_k=-1$ as $T\to \infty $. 
  Both cases coincide
with the results of the model on a static lattice 
(see , e.g.,\cite{a13,t2}).

\section{ Free energy on the Torus at Arbitrary $T$ }
\indent

 One can reduce the $2N^2$ integrals
over $X_{ij} , Y_{ij} $ to $2N$ integrals over their eigenvalues
$x_i\, , \, y_i$ obtaining \cite{a9}\cite{a10} :
\be
E(g,c,H)\, = \, -\frac{1}{N^2}\, {\rm ln}\, 
           \frac{Z(g,c,H)}{Z(0,c,H)}  \, ,                 \label{eq:r18}
\ee
where 
\be
Z(g,c,H)\, =\, \int_{-\infty}^{\infty}\int_{-\infty}^{\infty}\,  
               \prod_{i=1}^N\, dx_i\, dy_i \,\omega_i(c,g,H)
               \prod_{k<j}(x_k-x_j)(y_k-y_j)              \label{eq:r19}
\ee                   
with
\be 
\omega_i \, =\, {\rm exp} \left\{ -\left( x_i^2 + y_i^2-2cx_iy_i + 
     \frac{ge^H}{N}x_i^4 + \frac{ge^{-H}}{N} y_i^4 \right) \right\}
                                                       \, . \label{eq:r20}
\ee
  We can further decrease the number of degrees of 
freedom introducing two sets of monic polynomials \cite{a9},
\be
P_j(x) \,  =  \, x^j + \sum_{k=0}^{j-1} a_k\, x^k         \label{eq:r21}
\ee
and
\be
Q_j(y) \,  = \, y^j + \sum_{k=0}^{j-1} b_k\, y^k \, ,    \label{eq:s21}  
\ee
which are orthogonal with respect to the weight eq.(\ref{eq:r20}),
\be
\int\int dx \, dy\ \omega (c,g,H) \, P_i(x)Q_j(y) \, =\, 
           \delta_{ij}\, h_i (c,g,H) \, .                \label{eq:r22}
\ee
Using the relations
\be 
\prod_{k<l} (x_k -x_l)(y_k - y_l) = 
             {\rm det} \, x_i^{j-1} \, {\rm det}\, y_i^{j-1} =
  {\rm det}\, P_{j-1}(x_i) \, {\rm det}\, Q_{j-1}(y_i)\, , \label{eq:s22} 
\ee
 one derives:
\bea
E (g,c,H)\, &=&\, - \frac{1}{N^2} \sum_{i=0}^{N-1}\, 
      {\rm ln}\, \frac{h_i(g,c,H)}{h_i(0,c,H)}         \\    \label{eq:r23}
 &=&\,  - \frac{1}{N^2} \sum_{i=1}^{N}\, (N-i)\, 
         {\rm ln}\, \frac{f_i(g,c,H)}{f_i(0,c,H)}
   -\frac{1}{N}\, {\rm ln}\, \frac{h_0(g,c,H)}{h_0(0,c,H)}\, , \label{eq:s23}
\eea
where $f_i \, =\, h_i/h_{i-1}\,$. 
    The last term of eq.(\ref{eq:s23}) can be calculated from
eq.(\ref{eq:r22}):
\bea
h_0 (g,c,H) \, &=& \, \int_{-\infty}^{\infty}\int_{-\infty}^{\infty}dx\, 
            dy\, e^{-\left( x^2+y^2-2c xy + \frac{ge^H}{N}x^4 +
             \frac{ge^{-H}}{N}y^4\right)} \\                 \label{eq:r24}
&=&\, h_0(0,c,H)\, \left[ 1 - \frac{1}{N}
                      \frac{3g \, {\rm cosh}(H)}{2(1-c^2)^2} +
       {\cal O}(\frac{1}{N^2}) \right] \,.                    \label{eq:s24}
\eea
The ratios $f_i$ and the auxiliary quantities 
$r_i, s_i, q_i$ and $t_i$ defined below,
\bea
xP_i(x)\, & = & \, P_{i+1}(x) +r_i P_{i-1}(x) + s_i P_{i-3}(x)+\cdots \\
                                                            \label{eq:r25}
yQ_i(y)\, & = & \, Q_{i+1}(y) +q_i Q_{i-1}(x) + t_i Q_{i-3}(y)+\cdots 
                                                            \label{eq:s25}
\eea
can be determined altogether by solving a set of coupled equations obtained
from integrals of total derivatives involving eq.(\ref{eq:r22}). 
   Since we are going to look at $\, N\to \infty \, $ 
it is more useful to present the continuum version of those equations. 
Introducing the notation:
\bea
x   \,   & = & \, i/N\, , \quad \quad \epsilon \, = \, 1/N\, , \quad \quad 
f_{i+k}\,  =  \, Nf(x+k\epsilon)\,, \\                      \label{eq:r26}
q_{i+k}\,& = &\, Nq(x+k\epsilon)\,,  \quad \quad r_{i+k} =
                                    \, Nr(x+k\epsilon)\,, \\  \label{eq:s26}
s_{i+k}\, &=&\, N^2s(x+k\epsilon)\,,  \quad 
\quad t_{i+k}=\, N^2t(x+k\epsilon)\,,                         \label{eq:t26}
\eea
the continuum equations become \cite{a4,a9}:
\be
cq(x)\, = \, f(x)\left\{ 1+ 2ug\left\lbrack r(x+\epsilon )+r(x) + 
             r(x-\epsilon) \right\rbrack \right\}\,,         \label{eq:r27}
\ee
\begin{eqnarray}
cf(x)+\frac{x}{2} - r(x)\,\, & = &  \,\,
           2ug \left\{ s(x+2\epsilon ) + s(x+\epsilon) + s(x) 
               \right.                                      \nonumber\\ 
         &  & \left.  \mbox{}  + r(x)\left\lbrack r(x+\epsilon )+
                    r(x)+r(x-\epsilon)\right\rbrack\right\}\,,  \label{eq:s27}
\end{eqnarray}
\be 
cs(x)\, =\, \frac{2g}{u} f(x) f(x-2\epsilon) f(x-\epsilon)  \label{eq:r28}
\ee
and three more
equations obtained from (\ref{eq:r27}-\ref{eq:r28}) by exchanging 
 $(u,r,s,q,t)\to (1/u,q,t,r,s)$, where $\, u = e^H$. 
    Using the expansion 
\be
f(x)\, = \, f_0(x) + \frac{1}{N} f_{1/2}(x) +\frac{1}{N^2}f_1(x)+\cdots
                                                            \label{eq:r29}
\ee
and analogous ones for $r(x),q(x),s(x),t(x)$ and their conjugates,
we reproduce, collecting the $(1/N)^0$ terms, 
the planar equation of \cite{a4}:
\be 
gx\, = \, \frac{c^2 z^3}{9} + \frac{z}{3}\left\lbrack \frac{1}{(1-z)^2}
-c^2\, +\frac{Bz}{(1-z^2)^2}\right\rbrack \equiv \, g(z) \,.   \label{eq:r30}
\ee
  Here $\,B=2\left\lbrack {\rm cosh}(H)-1\right\rbrack \, $
and $\, z(x)=(6g/c)f_0(x) \, $. The above equation furnishes $\, f_0(x)\,$
parametrically. By further collecting the $\, (1/N) \, $ terms
we get \\ 
\parbox{13cm}{\begin{eqnarray*}
f_{1/2}\,\, & = & \,\, q_{1/2} \, =\, r_{1/2} \, = \, 0 \, ,          \\
t_{1/2} \,\, & = & \,\, -\frac{dt_0}{dx}\, ,\makebox[0.6cm]{} s_{1/2} \, = \, 
              - \frac{ds_0}{dx} \, . 
\end{eqnarray*}}\hfill\parbox{1cm}{\begin{eqnarray}\label{eq:r31}
\end{eqnarray}} \\
Using such results we derive from the $\, (1/N)^2 \, $ terms:

\begin{eqnarray}
\lefteqn{f_1 \bigl(z(x)\bigr) }                \nonumber\\ 
& & =\, \frac{c\,g}{2}\,z\,\bigl( 1 - z^2 \bigr)^4 \,  
           \left\lbrace 4(1+z)^8 + 2B(B+4)
           \left\lbrack(1+z^2)^4 + 32z^4 \right\rbrack   
                                         \right.         \nonumber \\
& & \makebox[0.5cm]{}  +\,  2\,{B^2} + 3\,{c^2} + c^4 + 
            16\,B\,z + 28\,{c^2}\,z +
            14\,B\,{c^2}z  + 75\,{c^2}\,{z^2}            \nonumber \\ 
& & \makebox[0.5cm]{}  -\, 5\,{c^4}\,{z^2} + 112\,B\,{z^3} + 
            16\,{c^2}\,{z^3} + 
            8\,B\,{c^2}\,{z^3} - 233\,{c^2}\,{z^4}       \nonumber \\ 
& & \makebox[0.5cm]{}  +\, 4\,{c^4}\,{z^4}  + 112\,B\,{z^5} - 
            308\,{c^2}\,{z^5} - 
            154\,B\,{c^2}\,{z^5} + 127\,{c^2}\,{z^6}     \nonumber \\
& & \makebox[0.5cm]{}  +\, 28\,{c^4}\,{z^6}  + 16\,B\,{z^7} + 
            512\,{c^2}\,{z^7} + 256\,B\,{c^2}\,{z^7} +  
            217\,{c^2}\,{z^8}                            \nonumber \\  
& & \makebox[0.5cm]{}  -\, 98\,{c^4}\,{z^8} - 268\,{c^2}\,{z^9} - 
            134\,B\,{c^2}\,{z^9} - 
            271\,{c^2}\,{z^{10}} + 154\,{c^4}\,{z^{10}}    \nonumber\\ 
& & \makebox[0.5cm]{}  -\, 16\,{c^2}\,{z^{11}} - 
            8\,B\,{c^2}\,{z^{11}} + 77\,{c^2}\,{z^{12}} - 
            140\,{c^4}\,{z^{12}} + 36\,{c^2}\,{z^{13}}     \nonumber\\ 
& & \makebox[0.5cm]{} +\, 18\,B\,{c^2}\,{z^{13}} +   
            5\,{c^2}\,{z^{14}} + \left. 76\,{c^4}\,z^{14} - 
            23\,c^4\,z^{16} + 3\,{c^4}\,z^{18}  \right\rbrace   \nonumber\\ 
& & \makebox[1.1cm]{}  \, \times \, 
            \left\lbrace (1+z)^4 \left\lbrack 1-c^2\,(1-z)^4 \right\rbrack 
           + 2\,B\,z\,(1+z^2)\right\rbrace^{-4}\, .           \label{eq:r32}
\end{eqnarray}
\bigskip
Another contribution of order $\, (1/N)^2 \, $ comes 
from applying the Euler-Maclaurin summation formula :
\be
\frac{1}{N} \sum_{i=1}^N F(i/N) =
\int_0^1 dx\,F(x) + \frac{1}{2N}\,[F(0)+F(1)] + \frac{1}{12N^2}
    F'(x)\,  \Big|_0^1 + {\cal O}(N^{-3})\, ,           \label{eq:r33}
\ee
on the expression (\ref{eq:r23}) which gives 
\bea
E (g,c,H)\, &=&  - \frac{1}{N}\, {\rm ln}\,h_0(g,c,H) -  
              \int_0^1 dx\,F(x)\,dx\, - \frac{1}{2N}\,[F(0)+F(1)]
                                                          \nonumber   \\
            & & \mbox{} - \frac{1}{12N^2}F^{'}(x)\,\Big|_0^1\,
              + {\cal O}(N^{-3})\, ,                    \label{eq:r34}
\eea
where
\be
F(x) = (1-x)\ {\rm ln}\left\lbrack \frac{f(x)}{f(x,g=0)} 
                                    \right\rbrack \, .   \label{eq:r35}
\ee

After using (\ref{eq:r30}), we obtain \\
\parbox{14cm}{\begin{eqnarray*}
g(z(x=1))\,~ & = & \,~ g  \\                               
g(z(x=0))\,~ & = & \,~ 0  \\                               
1-x      \,~ & = & \,~ 1 - \frac{g(z)}{g}\, .                       
\end{eqnarray*}}\hfill\parbox{1cm}{\begin{eqnarray}\label{eq:r36}
\end{eqnarray}} \\

The integral in (\ref{eq:r34}) can be written, after an 
integration by parts and using (44) , as 
\be
I(g,c,H) \equiv \int_0^1 dx\,(1-x)\, {\rm ln}\,f(x)  =
 I_0(g,c,H) + \frac{1}{N^2}\, I_1(g,c,H) + {\cal O}(N^{-4})\, , 
                                                            \label{eq:r40}
\ee
where $I_0$ is the planar contribution:
\be
I_0(g,c,H) = -\frac{1}{2}\left\lbrack (1-x)^2\,{\rm ln}\,f(x)  \right\rbrack 
            {\Big|_0^1} + \frac{1}{2g^2}
       \int_{z(0)}^{z(1)}\frac{dz'}{z'}\,[g-g(z')]^2 \, ,      \label{eq:r41}
\ee
while $I_1$ stands for the torus contribution,
\be
I_1(g,c,H) = -\frac{1}{2}\,(1-x)^2\,\frac{f_1(x)}{f_0(x)} +
 \frac{3}{gc}\int_{z(0)}^{z(1)} dz'\, \,[g-g(z')]^2\,\frac{d}{dz'}
\left\lbrack \frac{f_1(z')}{z'} \right\rbrack \, .           \label{eq:r42}
\ee

 In the above formulas $z(1)$ is the solution of eq.(\ref{eq:r30})
with $x=1$ that vanishes for $g=0$ 
and $z(0)=0$ since it is the only solution 
of eq.(\ref{eq:r30}) for $x=0$ that is real 
in the whole range of temperatures.
After further manipulations we arrive at 
\be
E(g,c,H) = E^{(0)}(g,c,H) + \frac{1}{N^2} E^{(1)}(g,c,H) + {\cal O}(N^{-4}) 
\, ,                                                          \label{eq:r43}
\ee
where
\be
E^{(0)}\bigl(g,c,B\bigr)=-\frac{1}{2}\, {\rm ln} \frac{z}{g}
+\frac{1}{g}\,\int_0^{z}\,\frac{dt}{t}\,g(t)
-\frac{1}{2 g^2}\,\int_0^{z}\,\frac{dt}{t}\,g^2(t)
-\frac{1}{2}\,{\rm ln} \Biggl(\frac{1-c^2}{3}e^{3/2}\Biggr)\, . \label{eq:r44}
\ee
Where , following the notation of [4] , the integration limit 
$z \equiv z(x=1)$ is of course a function of
$g, c$ and $B$.
This is the known result from \cite{a4} for the free energy of the 
Ising on random planar lattices (sphere).
The next term which is not known in the literature 
is the torus contribution, \\
\parbox{14cm}{\begin{eqnarray*}
E^{(1)}\bigl(g,c,B\bigr)\,  & = &\,  g\,\frac{{\rm cosh}(H)}{(1-c^2)^2}
+\frac{1}{12}\,{\rm ln}\, \biggl(\frac{1-c^2}{3}\biggr) 
         +\frac{1}{12}\ln\,\frac{z}{g}-3\,g\, R(z)        \\
\, &   & \, + 6\,\int_0^{z}\,dt\,g(t)\,\frac{d~ R(t)}{d~ t} -
         \frac{3}{g}\,\int_0^{z}\,dt\,g^2(t)\,\frac{d~ R(t)}{d~ t}\, . 
\end{eqnarray*}}\hfill\parbox{1cm}{\begin{eqnarray}\label{eq:r45}
\end{eqnarray}} \\
Here we have define 
$ R(z) \equiv f_1(z)/{cz} $.
  Note that the $1/N$ contribution vanishes in (\ref{eq:r43}),
which is in agreement with the expected topological expansion \cite{b1}.

\section{Yang-Lee Zeros on the Torus}
\indent

 The free energies $E^{(0)}$ and $E^{(1)}$ depend on
the coupling $g$ explicitly and implicitly through $z$.
 Now we can expand $E^{(1)}$ around $g=0$ using
\be
z(x=1,g) = \sum_{k=1}^{\infty}\, \frac{1}{k!}\,h_k\, g^k\, , \label{eq:r46}  
\ee

\noindent where $h_k = h_k(c,B)$ can be calculated from consecutive derivatives
of $g(z)=g$ with respect to $g$ at the point $g=0$. Moreover, the
integral terms in eq.(\ref{eq:r45}) may be combined according to

\be
\Lambda(t)\equiv \Bigl[g^2(t)-2g\,g(t)\Bigr]\,\frac{d~ R(t)}{d~ t},
                                                             \label{eq:r47}  
\ee
and the resulting integral expanded in powers of $z$, 

\be
\int_0^{z}\,dt\,\Lambda(t)=\sum_{k=0}^{\infty}\, \frac{1}{(k+1)!}\,
z^{k+1}\,\Bigl[\partial^{k}_t\,\Lambda(t)\Bigr]_{t=0}.       \label{eq:r48}  
\ee
In eq.(\ref{eq:r48}), $z$ is given by the series in eq.(\ref{eq:r46}),
and $\Bigl[\partial^{k}_t\,\Lambda(t)\Bigr]_{t=0}$
is a function of $g$, $c$ and $B$.

Using eqs.(\ref{eq:r46}) and (\ref{eq:r48}) together with the 
expansion of $R(z)$ in powers of $g$, and comparing with 
eq. (\ref{eq:r8}), we obtain the partition functions on 
random toroidal lattices $Z_{n}^{(1)}$
up to $n=16$ vertices . 
The first 8 results are collected in Table 1.

\begin{table}[ht]
\begin{tabular}{ll}
\\[-0.45cm]
\hline
\\[-0.4cm]
$Z^{(1)}_1 =$  & $\frac{1}{4}\, c^{-1}\,y^{-1/2}\,\left(1 + y \right) $    \\
         &          \\
$Z^{(1)}_2 =$  & $\frac{5}{8}\, c^{-2}y^{-1}\,\left[ 3 + (4\,{c^2} + 2\,{c^4})\,y  
                       +\: 3 {y^2} \right] $                          \\
         &          \\
$Z^{(1)}_3 =$  & $\frac{3}{2}\, c^{-3}\,y^{-3/2}\,\left( 1 + y \right) \,
                \left[ 11 
                    +\: (- 11 + 16\,{c^2} + 17\,{c^4})\,y  
                    +\: 11\,{y^2} \right]  $                              \\
         &          \\    
$Z^{(1)}_4 =$  & $\frac{3}{16}\, c^{-4}\,y^{-2} \left[ 837 
                    +\: (1416\,{c^2} + 1932\,{c^4})\,(y + y^3) 
                    +\: (936\,{c^2} + 2708\,{c^4}               \right. $ \\
         & $ \left. +\, 1304\,{c^6} +\: 74\,{c^8})\,{y^2} 
                    +\: 837\,{y^4}                             \right]  $ \\
         &          \\
$Z^{(1)}_5 =$  & $\frac{27}{10} \,{c^{-5}}\,{y^{-5/2}} \left( 1 + y \right) \,
              \left[ 579 +\: (-579 + 1140\,{c^2} + 1755\,{c^4})\,(y + y^3)  
                                                                \right. $ \\
         & $ \left. +\: (579 - 525\,{c^2} + 805\,{c^4} + 2245\,{c^6} 
                    + 370\,{c^8})\,{y^2} +\: 579\,{y^4}\right]    $ \\
         &          \\
$Z^{(1)}_6 =$  & $\frac{3}{4} {c^{-6}}\,{y^{-3}} \,\left[ 21411 +\, (48492\,{c^2} + 
              79974\,{c^4})\,(y + y^5) 
             +\, (23382\,{c^2} + 119160\,{c^4}                 \right. $ \\ 
          & $ \left. +\, 142434\,{c^6} + 36189\,{c^8})\,(y^2 + y^4)  
                     +\: (18630\,{c^2} + 115452\,{c^4}  
                     + 188488\,{c^6}                            \right. $ \\ 
          & $ \left. +\, 92868\,{c^8} + 12522\,{c^{10}}  
                     + 260\,{c^{12}})\,{y^3} +\, 21411\,{y^6}   \right]  $ \\
          &          \\
$Z^{(1)}_7 =$  & $\frac{81}{7} \,{c^{-7}}\,{y^{-7/2}} \left( 1 + y    \right) \,
            \left[    14571 +\: (-14571 + 37464\,{c^2} 
                    +  64533\,{c^4})\,(y + y^5)                 \right. $ \\ 
         & $ \left. +\: (14571 - 20580\,{c^2} + 34076\,{c^4}  
                    +  143444\,{c^6} + 47054\,{c^8})\,(y^2 + y^4) 
                                                                \right. $ \\  
         & $ \left. +\: (-14571 + 32319\,{c^2} + 56581\,{c^4} 
                    +  61278\,{c^6} + 111853\,{c^8} + 41307\,{c^{10}} 
                                                               \right. $ \\  
         & $ \left. +\, 2653\,{c^{12}})\,{y^3} +\, 14571\,{y^6}  \right] $ \\
         &          \\
$Z^{(1)}_8 =$ & $\frac{81}{32}\,{c^{-8}}\,{y^{-4}} \left[ 710991  
                    +\: (2050416\,{c^2} + 3637512\,{c^4})\,(y + y^7) 
                    +\: (884736\,{c^2}                         \right. $  \\
         & $ \left. +\,  5729832\,{c^4} + 9564480\,{c^6} 
                    +  3728700\,{c^8})\,(y^2 + y^6) +\, (564192\,{c^2} 
                                                               \right. $  \\
         & $ \left. +\, 5038200\,{c^4} + 14023200\,{c^6}  
                    +   14328000\,{c^8} + 5312016\,{c^{10}}     \right. $  \\
         & $ \left. +\, 549888\,{c^{12}})\,(y^3 + y^5)        
                    +\: (487872\,{c^2} + 4742568\,{c^4} 
                    +   14910912\,{c^6}                         \right. $  \\
         & $ \left. +\, 18385900\,{c^8} + 9308480\,{c^{10}} 
                    +   1819608\,{c^{12}} + 112832\,{c^{14}}  
                    +   1198\,{c^{16}})\,{y^4}                  \right. $ \\ 
         & $ \left. +\: 710991\,{y^8} \right]                           $ \\ 
         &  \\
\hline
\end{tabular}
\vspace{0.1cm}
\caption{\baselineskip=0.8cm The first eight partition functions 
            on Graphs with torus topology.}
\end{table}

\begin{figure}[ht]
\renewcommand{\captionlabeldelim}{.~}
\renewcommand{\figurename}{Fig.}
\begin{center}
\begin{minipage}[t]{1.0\textwidth}
\centering
\includegraphics[angle=-90,width=0.72\textwidth]{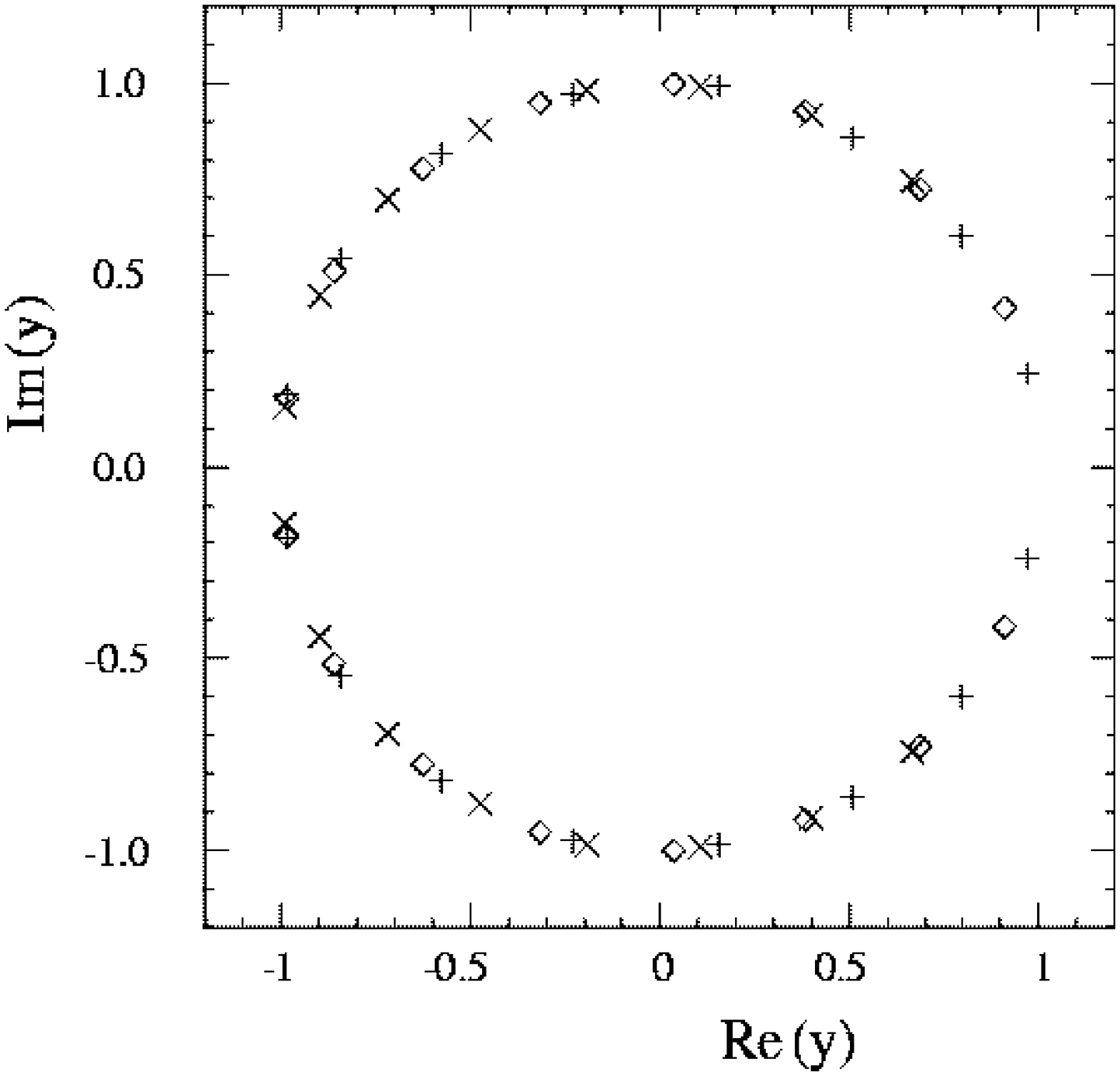}
\caption{Yang-Lee zeros of $Z^{(1)}_{16}$ on torus, for temperatures 
         $c= 0.25 (+),\, 0.45 (\diamond)\, ~{\rm and}\,~ 0.65 (\times)$.}
\label{fig4}
\end{minipage}
\end{center}
\end{figure}
\begin{figure}[ht]
\renewcommand{\captionlabeldelim}{.~}
\renewcommand{\figurename}{Fig.}
\begin{center}
\begin{minipage}[t]{1.0\textwidth}
\centering
\includegraphics[angle=-90,width=0.72\textwidth]{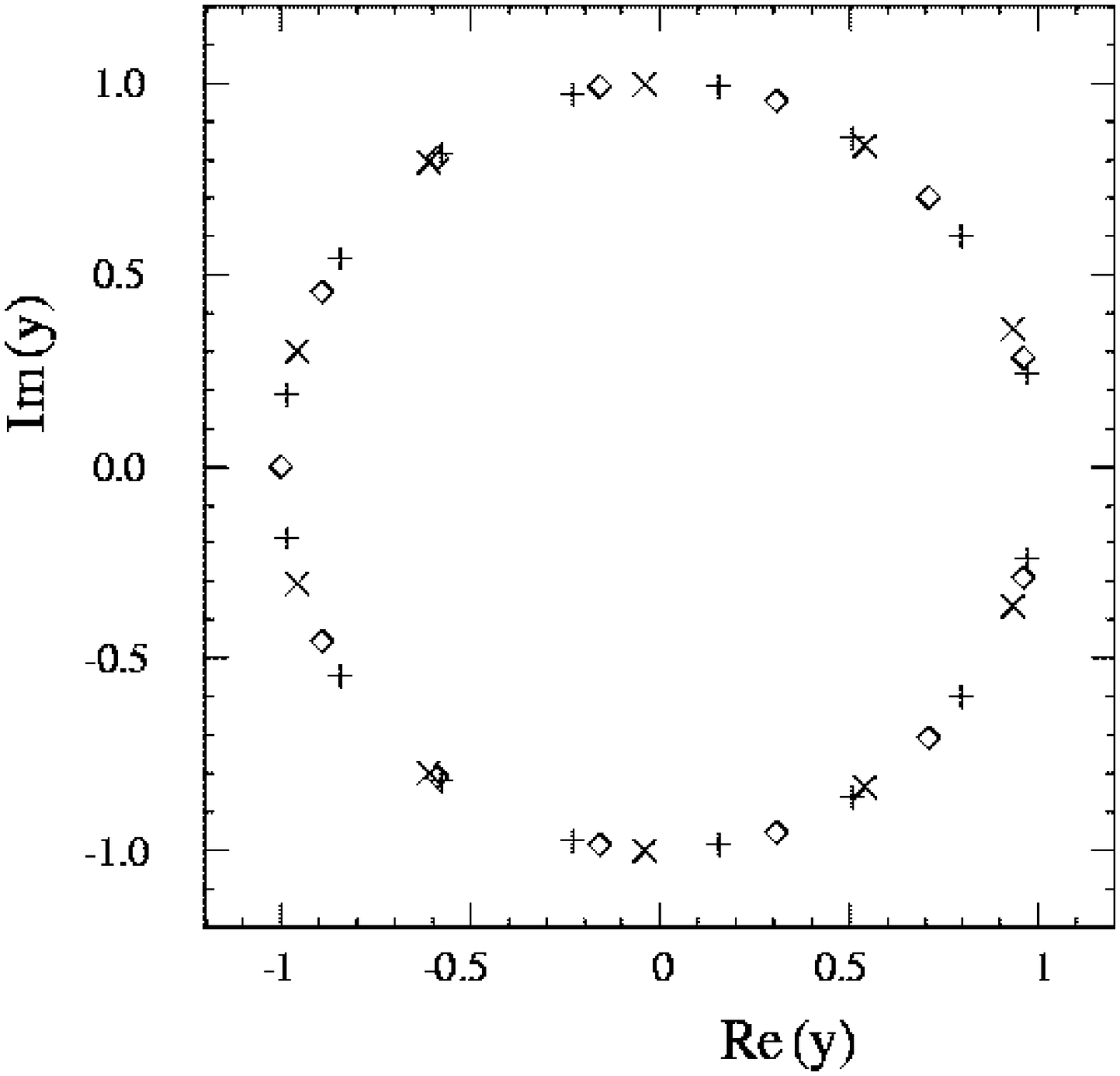}
\caption{Yang-Lee zeros of $Z^{(1)}_{n}$  
         at critical temperature $c=0.25$ for
         different lattice sizes $n$ on torus: 
         $n=10 (\times), 13 (\diamond)\, ~{\rm and}\,~ 16 (+)$.}
\label{fig5}
\end{minipage}
\end{center}
\end{figure}
The figure 4 shows the flux of zeros on the unit circle, as the 
temperature $c$ decreases to the critical one, $c = 0.25$.
In Fig. 5 we clearly see that the zeros tend to pinch 
the positive real
$y-$axis as the number of sites increases 
(thermodynamic limit) at the critical temperature
$c = 0.25$. 
It is worth mentioning that the polynomials are different from their 
planar counterparts, but their zeros still lie on the unit circle.
The topology slightly changes the position of the zeros along the
unit circle.


\begin{table}[t]
\begin{tabular}{ll}
\\[-0.45cm]
\hline
\\[-0.4cm]
$Z_1 =$  & $\frac{3}{4}\,c^{-1}\,y^{-1/2}\, \left(1 + y \right) $ \\
         &  \\
$Z_2 =$  & $ \frac{3}{2}\, c^{-2}\,y^{-1}\, 
            \left[ 2 +\: (3\,c^2 + c^4)\,y +\: 2\,y^2 \right] $ \\
         &  \\
$Z_3 =$  & $\frac{9}{4}\,c^{-3}\,y^{-3/2}\, (1 + y)\,
            \left[ 11 
           +\: (-11 + 16\,c^2 + 17\,c^4)\,y +\: 11\,y^2 \right] $ \\
         &  \\
$Z_4 =$  & $\frac{9}{4}\, c^{-4}\, y^{-2}\,
            \left[ 136 + (198\,c^2 + 346\,c^4)\,(y\,+\,y^3) 
           +\: (128\,c^2 + 429\,c^4 +       \right.              $ \\
         & $ \left.  242\,c^6 + 17\,c^8)\,y^2   +\: 136\,y^4 \right] $ \\
         &  \\
$Z_5 =$  & $\frac{27}{20}\,c^{-5}\,y^{-5/2}\,
            (1 + y) \left[3714 
           +\: (-3714 + 5440\,c^2 + 13130\,c^4)\,(y\,+\,y^3) \right.  $ \\
         & $ \left.  +\: (3714 - 
              2800\,c^2 + 1455\,c^4 + 16370\,c^6 + 3545\,c^8)\,y^2  
            +\: 3714\,y^4 \right] $ \\
         &  \\
$Z_6 =$  & $\frac{9}{2}\,c^{-6}\,y^{-3}
             \left[22688 + (33426\,c^2 + 102702\,c^4)\,(y\,+\,y^5) +\: 
             (13056\,c^2 + 99855\,c^4 \right.      $ \\
         & $ \left. + 167082\,c^6 + 60327\,c^8)\,(y^2\,+\, y^4) 
             +\: (9801\,c^2 + 89211\,c^4 + 199254\,c^6 \right.  $ \\
         & $ \left. + 131184\,c^8 + 23601\,c^{10}  
              + 709\,c^{12})\,y^3 +\: 22688\,y^6 \right] $ \\
         &  \\
\hline
\end{tabular}
\vspace{0.1cm}
\caption{\baselineskip=0.8cm The first six
all topologies  partition functions $Z_n$ for $N=1$. } 
\end{table}

\begin{table}[ht]
\begin{tabular}{ll}
\\[-0.45cm]
\hline
\\[-0.4cm]
$Z_1 =$ & $ \frac{9}{16}\,c^{-1}\,y^{-1/2}\:\left(1 + y      \right) $ \\
        &  \\
$Z_2 =$ & $ \frac{3}{32}\,c^{-2}\,y^{-1}\:
            \left[  17 + (28\,c^2 + 6\,c^4)\,y +\: 17\,y^2   \right]  $ \\
        &  \\
$Z_3 =$ & $ \frac{81}{64}\,c^{-3}\,y^{-3/2}\:
             (1 + y)\: \left[ 7 +\: (-7 + 12\,c^2 + 9\,c^4)\,y 
            +\: 7\,y^2                                      \right]  $ \\
        &  \\
$Z_4 =$ & $ \frac{9}{256}\,c^{-4}\,y^{-2}\:
             \left[  2011 + (3648\,c^2 + 4396\,c^4)\,(y\,+\,y^3) 
            +\: (2528\,c^2 + 6424\,c^4                       \right.  $ \\
        & $  \left. +\: 2952\,c^6 + 162\,c^8)\,y^2  
            +\: 2011\,y^4                                    \right]  $ \\
        &  \\        
$Z_5 =$ & $ \frac{81}{1280}\,c^{-5}\,y^{-5/2}\:
            (1 + y)\:\left[ 11394 +\: (-11394 + 21740\,c^2 +
            35230\,c^4)\,(y\,+\,y^3)                          \right.  $ \\
        & $ \left.  + \:
            (11394 - 9800\,c^2 + 14055\,c^4 + 44770\,c^6 + 
            7945\,c^8)\,y^2  +\: 11394\,y^4                  \right]  $ \\
        &  \\
$Z_6 =$ & $ \frac{9}{256}\,c^{-6}\, y^{-3}\:
            \left[ 252169 +\: (501588\,c^2 + 1011426\,c^4)\,(y\,+\,y^5) +\: 
             (231528\,c^2                                    \right.  $ \\
        & $ \left.  +  1299915\,c^4 
              +  1737966\,c^6 + 513126\,c^8)\,(y^2\,+\,y^4) 
              + \: (183438\,c^2 + 1229793\,c^4              \right.  $ \\
        & $ \left.  +  2214752\,c^6 + 1222392\,c^8 + 
             188238\,c^{10} + 4767\,c^{12})\,y^3 + \:
             252169\,y^6                                     \right]  $ \\
        &  \\
\hline
\end{tabular}
\vspace{0.1cm}
\caption{\baselineskip=0.8cm The first six all topologies 
partition  functions for $N=2$.}
\end{table}

\section{Non-perturbative Partition Function for Small Matrices}
\indent

  As we have seen in section 3 the exact free energy for spherical
and torus contributions were given as integral representations 
(eqs. (\ref{eq:r44}) and (\ref{eq:r45})), in the large $N$ limit. 
  These turn out to be the first two contributions of a topological
expansion in $N^{-2h}$ \cite{b1}, where $h$ stands for the
       number of handles of the corresponding surface in the 
continuum limit.
 On the other hand, if we consider small values for $N$, it is
possible to obtain an explicit closed form for the free energy
(\ref{eq:r18}) including all topologies which contribute
to  a given number of vertices.
Perturbatively in $g$ , we have in those cases few 
gaussian integrals to perform 
( see (\ref{eq:r19})).

 It is convenient to write  eq. (\ref{eq:r19}) as follows,
\be
Z(g,c,H)\, =\, Z(0,c,H)\, +\, \tilde{Z}(g,c,H)\,.          \label{eq:r49}
\ee                   
From eqs. (\ref{eq:r49}) and (\ref{eq:r18}) and, 
after expanding the logarithm, we obtain the free energy 
\be
E(g,c,H)\, = \, -\frac{1}{N^2}\, \sum_{m=1}^{\infty}
             \frac{(-1)^{m+1}}{m}\,
   \left[ \frac{\tilde{Z}(g,c,H)}{Z(0,c,H)} \right ]^m \, .   \label{eq:r50}
\ee
 The above expansion can be rewritten in powers of $g$, 
\be
\frac{\tilde{Z}(g,c,H)}{Z(0,c,H)} \, = \, \sum_{p=1}^{\infty}
                   z_N^{(p)}(c,H)\,\, g^p \, .                  \label{eq:r51}
\ee
The coefficients $ z_N^{(p)}$ can be easily calculated for
small number of eigenvalues. 
Here we restrict ourselves , for sake of simplicity , 
to $N=1$ (no contribution 
from the vandermonde ) and $N=2$
.After rearranging (55) in powers of the coupling $g$
and comparing with the expression (8) 
we obtain the nonperturbative partition functions $Z_n$ for $N=1,2$.
We have obtained $Z_n$ explicitly for
$n\le 20 $ sites .
 In tables 2 and 3 we display the 
nonperturbative partition functions 
$Z_n$\, $(n=1,2,..., 6)$ for $N=1$ and $2$, respectively.
Finally, in Fig. 6 we show the Yang-Lee zeros for our largest
lattice, $n=20$ sites, for $N=2$, which are quite close
to the ones for $N=1$. Here, we clearly see  
 the flux of zeros on the unit circle, as the 
temperature $c$ decreases to the critical one, $c = 0.25$.
\begin{figure}[ht]
\renewcommand{\captionlabeldelim}{.~}
\renewcommand{\figurename}{Fig.}
\begin{center}
\begin{minipage}[t]{1.0\textwidth}
\centering
\includegraphics[angle=-90,width=0.72\textwidth]{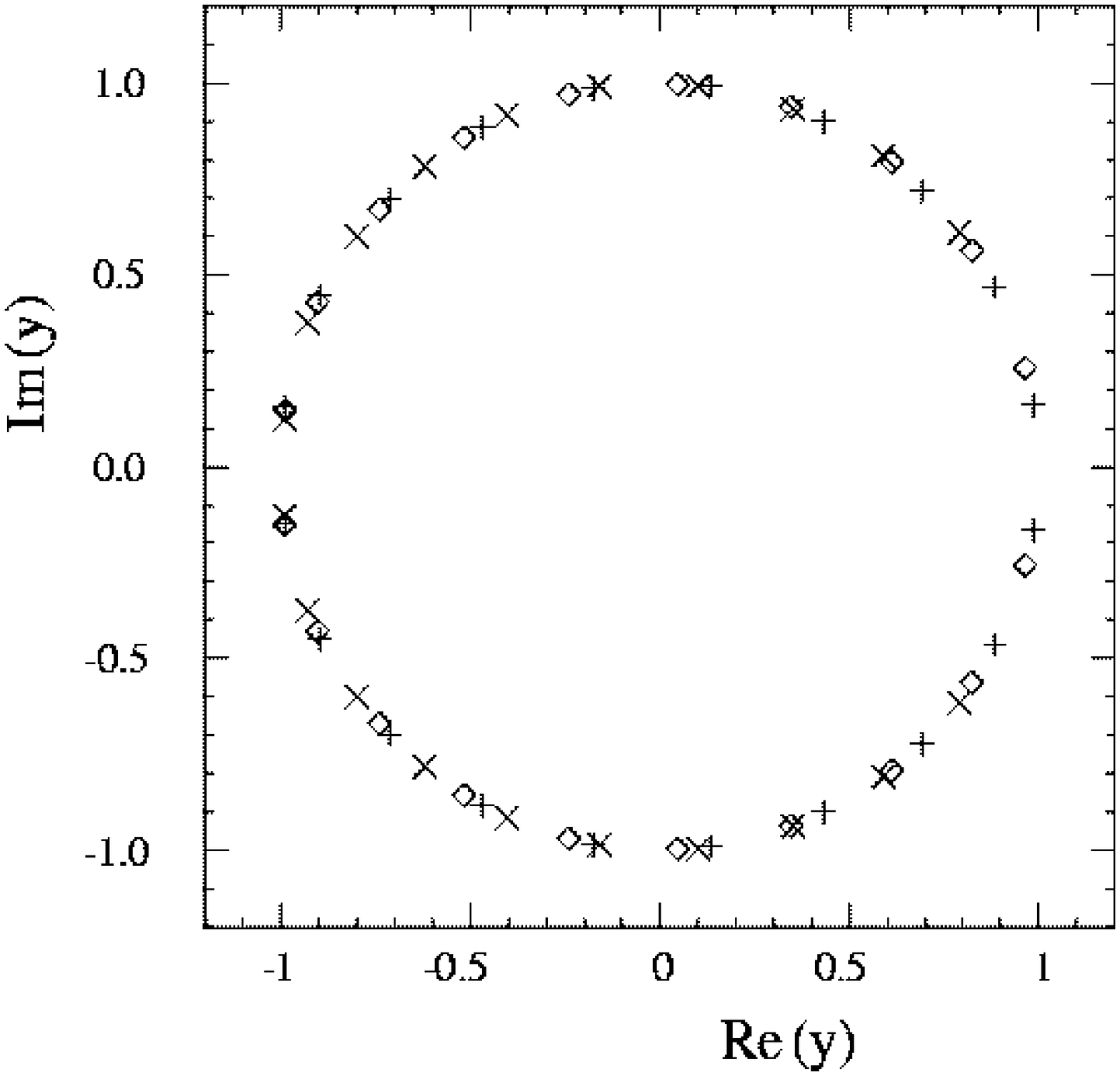}
\caption{
Yang-Lee zeros of $Z_{20}$ with $N=2$, for temperatures 
         $c= 0.25 (+),\, 0.45 (\diamond)\, ~{\rm and}\,~ 0.65 (\times)$.} 
\label{fig6}
\end{minipage}
\end{center}
\end{figure}

\section{Conclusion}
\indent

  We have obtained a closed and exact expression for
the $\, 1/N^2 \, $ contribution to the free energy of
the two Hermitian random matrix model. From this expression 
one can obtain the partition function $\, Z_n^{(1)} \, $
of the Ising model in the presence of a constant magnetic
field $H$  on random graphs (2D Gravity) of torus topology
with $\, n \,$ sites. 
  For arbitrary temperature we have 
obtained explicitly the partition functions
$\,Z_n^{(1)} \,$ with $n\le 16$.
For the specific temperatures $c=0.25,\,0.45$ and $0.65$, 
we have checked that the zeros of the polynomials 
$\,Z_n^{(1)} \,$ $(n\le 16)$ lie on the
unit circle $(\,\vert y \vert = 1)$ on the complex fugacity plane.

For the special cases of small matrices $ 1\times 1 $ and $ 2\times  2 $ 
we have obtained the non-perturbative  
partition function $Z_n$ which includes contributions of
all topologies. We have explicit results for
$n\le 20 $ sites and again the Yang-Lee zeros
lie on the unit circle.
That is quite surprisingly since there is
no Lee-Yang circle theorem for dynamical lattices. 
Taking linear combinations of polynomials in general originates 
non easily predictable changes in their roots. A similar result
has been observed before for the $\,Z_n^{(0)} \,$ $(n\le 14)$
on dynamical planar graphs. Our calculations on the torus 
and for higher topologies is thus a strong evidence that  
the topology of the graph plays no special role
what the position of the Yang-Lee zeros is concerned.
The same happens on a static lattice since
the Lee-Yang theorem (see \cite{a2})
 is known to be independent on
the details of the lattice like its topology, number of 
nearest neighbors, etc. 
It should be stressed that our dynamical lattice consists 
of sums of $\phi^4\,$ vacuum to vacuum diagrams where the weight of
each diagram corresponds to its combinatorial factor.
Nevertheless, those specific weights seem to be immaterial  
concerning the unit circle.
It is remarkable that replacing those specific weights by arbitrary 
positive constants, at least for $\, n\le 4\,$ sites, 
we still have the roots on the unit circle. 
That happens on the sphere, on the torus and also for linear 
combinations, with positive constants, of both topologies.
In fact this is probably the reason for 
the location on the unit circle of 
the zeros of the nonperturbative 
partition functions that we have 
obtained for the special cases of $1\times 1$
and $2\times 2$ matrices.
In those cases we have a linear combination
of all topologies wiht the positive coefficiens
$N^{-2h}$ where $h=0,1,\cdots \, \infty $ labels the different topologies . 
Such results lead us to conjecture that 
linear combinations , with positive 
coefficients ,  of partition functions of 
the 2D Ising model calculated on different graphs $G_n$ 
have all their $n$ Yang-Lee zeros on the unit circle.
At this point a word of caution is in order ,
namely , Lee and Yang have proven their
circle theorem for a class of polynomials
which includes the partition function
of the Ising Model in a constant magnetic field ,
nevertheless ,
it is easy to find numerical examples of 
linear combinations of such polynomials which
donot obey the circle theorem even if we 
take positive coefficients. 
 Therefore it is not
obvious how the Lee-Yang theorem can be generalized for
sums over the graphs (lattices).
It cannot be discarded that 
the polynomials we have for the dynamical case 
correspond to a new class of
polynomials not encompassed by the Lee-Yang theorem but whose
roots still lie on the unit circle. A clear understanding of this
issue demands more work which is now in progress.
We emphasize that we have only looked at the global location
of the Yang-Lee zeros rather than their local
distribuition close to the positive real axis 
and its relationship with the topology.
This is now under investigation.
Finally, it is worth pointing out that
we were able to sum over all topologies for a 
given finite number of vertices because the sum is
actually finite since not all topologies can contribute
to a graph with a finite number of vertices with finite
number of links. Thus , we have a natural cut-off .
\section{Note Added}
\indent

After finishing this work we became aware of \cite{dj}
where for the special case $N=1$ ("Thin Graphs")
it was shown in the thermodynamic limit 
that the closest zero to the positive real axis 
is located on the unit circle on the complex fugacity plane.
That is clearly in agreement with the expectations
from our finite size calculations  of  {\bf  all} 
Yang-Lee zeros in the case $N=1$ (see section 5).

\section{Acknowledgements}
\indent
D.D. is indebted to R.Schrock for an interesting discussion. 
This work was partially supported by FAPESP
under contracts \# 96/11747-2 , \# 97/06499-2 , \# 98/12626-0 (D.D.) 
and \# 97/01378-2 (L.C.A.). 
L.C.A. would like to thank the Departamento de Fisica Matem\'atica  
of Universidade de S\~ao Paulo for their hospitality.




\vskip 2.0cm
{\Large Figure Captions:}\\
Fig. 1.  {Diagrams and partition functions 
          for $n=2$ vertices.              \\
Fig. 2.  Diagrams and partition functions for $n=3$ vertices. \\
Fig. 3a. First set of diagrams and partition functions 
         for $n=4$ vertices.  \\
Fig. 3b. Second set of diagrams  and partition functions 
         for $n=4$ vertices. \\ 
Fig. 4.  Yang-Lee zeros of $Z^{(1)}_{16}$ on torus, for temperatures 
         $c= 0.25 (+),\, 0.45 (\diamond)\, ~{\rm and}\,~ 0.65 (\times)$.\\
Fig. 5. {Yang-Lee zeros of $Z^{(1)}_{n}$  
         at critical temperature $c=0.25$ for
         different lattice sizes $n$ on torus: 
         $n=10 (\times), 13 (\diamond)\, ~{\rm and}\,~ 16 (+)$.\\
Fig. 6.  Yang-Lee zeros of $Z_{20}$ with $N=2$, for temperatures 
         $c= 0.25 (+),\, 0.45 (\diamond)\, ~{\rm and}\,~ 0.65 (\times)$.} \\

\end{document}